\documentclass[12pt]{article}
\usepackage{graphicx}
\usepackage{cite}
\newcommand{\abs}[1]{\left| #1\right|}
\newcommand{\fnd}[2]{\frac{\textstyle #1}{\textstyle #2}}
\newcommand{\xrm}[1]{{\textstyle \mbox{\rm #1}}}
\newcommand{\bm}[1]{\mbox{\boldmath $#1$}}
\newcommand{\Real}[1]{\Re {\it e}(#1 )}
\newcommand{\Imag}[1]{\Im {\it m}(#1 )}
\newcommand{\fsml}[2]{\mbox{$\frac{#1}{#2}$}}
\newcommand{\dsb}{$D^{\ast}_{sJ}(2463)$}
\newcommand{\dsn}{$D_{s1}(2536)$}
\newcommand{\db}{$D_{1}(2400)$}
\newcommand{\dn}{$D_{1}(2420)$}
\begin{document} \baselineskip .7cm
\title{Continuum bound states
$K_{L}$, $D_{1}(2420)$, $D_{s1}(2536)$,\\ and their partners
$K_{S}$, $D_{1}(2400)$, $D^{\ast}_{sJ}(2463)$}
\author{
Eef van Beveren\\
{\normalsize\it Centro de F\'{\i}sica Te\'{o}rica}\\
{\normalsize\it Departamento de F\'{\i}sica, Universidade de Coimbra}\\
{\normalsize\it P-3000 Coimbra, Portugal}\\
{\small eef@teor.fis.uc.pt}\\ [.3cm]
\and
George Rupp\\
{\normalsize\it Centro de F\'{\i}sica das Interac\c{c}\~{o}es Fundamentais}\\
{\normalsize\it Instituto Superior T\'{e}cnico, Edif\'{\i}cio Ci\^{e}ncia}\\
{\normalsize\it P-1049-001 Lisboa Codex, Portugal}\\
{\small george@ajax.ist.utl.pt}\\ [.3cm]
{\small PACS number(s): 12.40.Yx, 14.40.Aq, 14.40.Lb, 13.25.Es, 13.25.Ft,
13.75.Lb}
}
\maketitle

\begin{abstract}
The very recently observed $D^{\ast}_{sJ}(2463)$ meson is described as a
$J^{P}\!=\!1^{+}$ $c\bar{s}$ bound state in a unitarised meson model,
owing its existence to the strong OZI-allowed $^{3}\!P_{0}$ coupling
to the nearby $S$-wave $D^{\ast}\!K$ threshold. By the same non-perturbative
mechanism, the narrow axial-vector $D_{s1}(2536)$ resonance shows up as a
quasi-bound-state partner embedded in the $D^{\ast}\!K$ continuum. With the
same model and parameters, it is also shown that the preliminary broad $1^{+}$
$D_{1}(2400)$ resonance and the established narrow $1^{+}$ $D_{1}(2420)$ may be
similar $c\bar{n}$ partners, as a result of the strong OZI-allowed
$^{3}\!P_{0}$ coupling to the nearby $S$-wave $D^{\ast}\!\pi$ threshold. The
continuum bound states $D_{1}$(2420) and $D_{s1}(2536)$ are found to be
mixtures of 33\% $^{3\!}P_{1}$ and 67\% $^{1\!}P_{1}$, whereas their partners
$D_{1}(2400)$ and $D^{\ast}_{sJ}(2463)$ have more or less the opposite
$^{2S+1}\!P_1$-state content, but additionally with some $D^{\ast}\!\pi$ or
$D^{\ast}\!K$ admixture, respectively.

The employed mechanism also reproduces the ratio of the $K_{L}$-$K_{S}$
mass difference and the $K_{S}$ width, by describing $K_{L}$ as a bound state
embedded in the $\pi\pi$ continuum.
The model's results for $J^{P}\!=\!1^{+}$ states containing one $b$ quark
are also discussed.
\end{abstract}

\section{Introduction}
Bound states embedded in the continuum have been suggested by von Neumann and
Wigner \cite{PZ30p465}. Since then many works on this theme have appeared in
various fields of physics
\cite{PRL90p013001,SAMARA97p435,GRQC9610018,QUANTPH9511022,HEPPH9305336}.
In the present paper, we shall study such states appearing as (approximate)
solutions of our simple unitarised meson model \cite{HEPPH0201006}. As three
concrete applications, we choose here the $K_L$-$K_S$ system, the \dsn\
\cite{PRD66p010001} together with the very recently observed narrow \dsb\
\cite{HEPEX0305100,HEPEX0305017}
state, and finally the couple consisting of the established
\dn\ \cite{PRD66p010001} and the preliminary broad \db\
\cite{BELLE02,CLEO01}
resonance. At first sight, it may seem odd to try to relate such utterly
disparate  states, ranging from mesons that can only decay weakly to very
broad mesonic resonances. Moreover, while in the $J^P\!=\!1^+$ $c\bar{n}$ ($n$
stands for non-strange) system the \db\ is much broader than the slightly
heavier \dn, for the $c\bar{s}$ states the likely $1^+$ \dsb\ is \em even
narrower \em \/than the $1^+$ \dsn. Nevertheless, we shall demonstrate below
that, in all three cases, a simple mechanism of coupling two degenerate
$q\bar{q}$ channels with the same quantum numbers to the meson-meson continuum
is capable of accounting for the experimental data, through an exact or
approximate decoupling of one of the physical $q\bar{q}$ states.

\section{Degeneracy lifting via coupling to the continuum}

When two degenerate discrete channels are coupled to the same continuum
channel, the degeneracy is lifted. One state decouples, partly or completely,
depending on the details of the interaction dynamics, and turns into a
continuum (approximate) bound state. The other state turns into either a
resonance structure in the continuum, or a bound state below threshold when
threshold is near enough. Within our unitarisation scheme, quark-antiquark
channels are coupled to the meson-meson continuum by OZI-allowed $^{3}\!P_{0}$
$q\bar{q}$ pair creation and annihilation.

In the present investigation, we confine our attention to two degenerate
quark-antiquark systems, which can be distinguished by some internal
structure irrelevant for the coupling to the meson-meson continuum.
For example, the constituent quark masses of $u$ and $d$ are the same in
most models.
Hence, the internal dynamics of $u\bar{u}$ and $d\bar{d}$ vector states
can in meson models be described by the same Hamiltonian.
Under $J^{PC}=0^{++}$ quark-pair creation, both systems couple with the same
intensity to pion pairs in $P$-wave.
Nevertheless, because of the relative sign of the coupling constants
under particle interchange \cite{ZPC21p291},
the $(u\bar{u}+d\bar{d})/\sqrt{2}$ state decouples from the two-pion
continuum ($\omega$ meson), whereas the $(u\bar{u}-d\bar{d})/\sqrt{2}$
state decays strongly into pion pairs ($\rho^{0}$ meson).

Let us denote by $\psi_{1}$ and $\psi_{2}$ the wave functions of the
quark-antiquark systems, and by $H_{1}$ and $H_{2}$ the Hamiltonians
describing their internal confinement dynamics ({\it e.g.} employing
a confinement potential).
For the continuum we write the wave function $\psi$ and the dynamics $H$.
The coupling interactions of the two confinement channels to the continuum
are denoted by $V_{1}$ and $V_{2}$.
In Ref. \cite{ZPC21p291} it is found how the full spin, isospin and color
degrees of freedom contribute to the determination of such potentials.
In the spirit of our model, we then obtain for the three channels
the following set of coupled dynamical equations
\cite{HEPPH0306155,CPC27p377,PRD27p1527,PRD21p772}.
\begin{eqnarray}
& (H-E)\;\psi\; +\; V_{1}\;\psi_{1}\; +\;V_{2}\;\psi_{2}\; =\; 0 &
\nonumber\\ [.3cm]
& V_{1}^{\dagger}\;\psi\; +\; (H_{1}-E)\;\psi_{1}\; =\; 0 &
\nonumber\\ [.3cm]
& V_{2}^{\dagger}\;\psi\; +\; (H_{2}-E)\;\psi_{2}\; =\; 0 &
\; .
\label{dynt}
\end{eqnarray}
We assume that the internal dynamics of the two degenerate
quark-antiquark systems does not depend on the difference in their internal
structure, and, moreover, that these systems also couple the same way to the
continuum, i.e.,
\begin{equation}
H_{1}\; =\; H_{2}
\;\;\;\;\;\;\xrm{and}\;\;\;\;\;\;
V_{1}\; =\;\alpha\; V
\;\;\; ,\;\;\;
V_{2}\; =\;\beta\; V
\;\;\;\;\;\;\;\; (\alpha^{2}+\beta^{2}=1) \; .
\label{V2propV1}
\end{equation}
In this case it is easy to sub-diagonalise the system (\ref{dynt}), so as to
obtain
\begin{equation}
\begin{array}{r}
(H-E)\;\psi\; +\; V\; \left(\alpha\psi_{1}+\beta\psi_{2}\right)\; =\; 0 \\[3mm]
V^{\dagger}\;\psi\;+\;(H_{1}-E)\;\left(\alpha\psi_{1}+\beta\psi_{2}\right)\;
=\; 0
\label{dyn12}
\end{array}
\end{equation}
and
\begin{equation}
(H_{1}-E)\;\left(\beta\psi_{1}-\alpha\psi_{2}\right)\; =\; 0
\;\;\; .
\label{dyn3}
\end{equation}
We end up with a system of the continuum $\psi$ coupled to a linear
combination $\alpha\psi_{1}+\beta\psi_{2}$ of the confinement states in
Eq.~(\ref{dyn12}), and with a completely decoupled system for the orthogonal
linear combination $\beta\psi_{1}-\alpha\psi_{2}$ in Eq.~(\ref{dyn3}).
Since, moreover, $H_{1}$ and $H_{2}$ are supposed to describe confinement,
Eq.~(\ref{dyn3}) has only bound-state solutions, all embedded in
the meson-meson continuum.

\section{The neutral kaon system}
\label{neutralK}

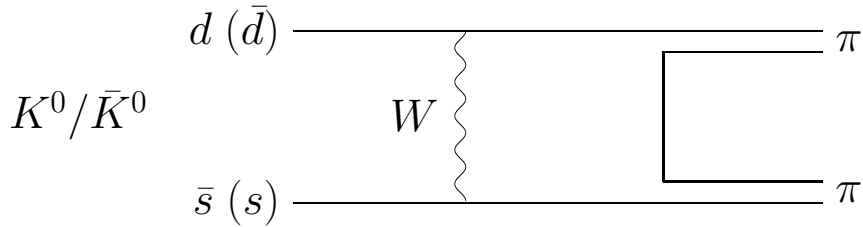
\begin{figure}[ht]
\Large
\begin{center}
\begin{picture}(260,80)(-120,-65)
\put(-60,0){\line(1,0){200}}
\put(80,-8){\line(1,0){60}}
\put(80,-57){\line(0,1){49}}
\put(80,-57){\line(1,0){60}}
\put(-65,0){\makebox(0,0)[rc]{$d\; (\bar{d})$}}
\put(145,-4){\makebox(0,0)[lc]{$\pi$}}
\put(-115,-32){\makebox(0,0)[rc]{$K^{0}/\bar{K}^{0}$}}
\put(-60,-65){\line(1,0){200}}
\put(-65,-65){\makebox(0,0)[rc]{$\bar{s}\; (s)$}}
\put(145,-61){\makebox(0,0)[lc]{$\pi$}}
\put(-5,-32){\makebox(0,0)[rc]{$W$}}
\end{picture}
\end{center}
\normalsize
\caption[]{$W$ exchange takes care of the flavour transitions triggering
the strong process of quark-pair creation in weak $K^{0}/\bar{K}^{0}$
two-pion decay.}
\label{Wexchange}
\end{figure}

A typical example of the above-described phenomenon is the two-pion decay mode
of the neutral kaon system.
Both $d\bar{s}$ and $s\bar{d}$ couple weakly to $\pi\pi$ via the process
depicted in Fig.~\ref{Wexchange}.
All other decay modes of neutral Kaons can to lowest order be neglected,
since they couple orders of magnitude weaker.

\noindent
Because of particle-antiparticle symmetry, we may assume
$\alpha =\beta =1/\sqrt{2}$.
The $CP\!=\!-1$ combination $\left(d\bar{s}\!-\!s\bar{d}\right)/\sqrt{2}$
completely decouples and turns into a bound state embedded in the
$\pi\pi$ continuum. This combination thus represents the $K_{L}$ meson,
which, since $CP$ violation is not contemplated in our model, has no
coupling to the two-pion continuum.
However, the $CP\!=\!+1$ combination
$\left(d\bar{s}\!+\!s\bar{d}\right)/\sqrt{2}$ exists as a, actually extremely
narrow, resonance in $S$-wave $\pi\pi$ scattering, hence describing the
$K_{S}$ meson.

We assume here that the decay process depicted in Fig.~\ref{Wexchange} is
dominated by the strong OZI-allowed $^3\!P_0$ quark-pair-creation mechanism,
whereas $W$ exchange merely functions as a trigger to this process,
which basically only determines the decay probability. Hence, besides the
smallness of the parity-violating $K_S\to\pi\pi$ coupling constant,
the transition potential describes here the coupling of a $u\bar{u}$ state with
$J^{PC}\!=\!0^{++}$, i.e., $\sigma$-meson quantum numbers to two pions, similar
to our unitarised description of scalar mesons studied in
Refs.~\cite{HEPPH0306155,EPJC22p493,HEPPH0110156}
(see also Ref.~\cite{MPLA14p1273}).
In the delta-shell approximation for the transition potential in the case of
the scalar $K_0^*(800)$ and $a_0(980)$ resonances
\cite{EPJC22p493,HEPPH0110156}, we obtained for the delta-shell radius $a$ a
value of about $a=3.2$ GeV$^{-1}$. We shall hold on to this value in the
following.

A general solution of Eq.~(\ref{dyn12}) for the $S$-wave scattering phase shift
$\xrm{cotg}\left(\delta (s)\right)$, as a function of the total invariant
two-pion mass $\sqrt{s}$, is in this approximation and for small coupling
given by
\begin{equation}
\xrm{cotg}\left(\delta (s)\right)\;\approx\;
\fnd{\left[ E_{0}\; +\; R(s)\;\abs{{\cal F}_{0}^{ds}}^{2}\right]
\; -\;\sqrt{s}}{I(s)\;\abs{{\cal F}_{0}}^{2}}
\;\;\; ,
\label{cotgdS}
\end{equation}
where $E_{0}$ represents the ground-state energy of the $d\bar{s}$
($s\bar{d}$) when uncoupled, hence the mass of the
$K_{L}$ meson.
The remaining factors $R(s)$, $I(s)$, and ${\cal F}_{0}$ are well explained in
Ref.~\cite{HEPPH0304105}.

\begin{figure}[ht]
\begin{center}
\begin{picture}(400,125)(0,-50)
\put(  0,-50){\line(1,0){400}}
\put(  0, 55){\line(1,0){400}}
\put(  0,-50){\line(0,1){105}}
\put(400,-50){\line(0,1){105}}
\put(0,0){\line(1,0){400}}
\put(100,-2){\line(0,1){4}}
\put(100,-2){\line(1,0){300}}
\put(100, 2){\line(1,0){300}}
\put(380,5){\makebox(0,0)[bl]{\small cut}}
\put(100,5){\makebox(0,0)[bl]{\small threshold}}
\put(100,-5){\makebox(0,0)[tc]{\small $\sqrt{s}=2m_{\pi}$}}
\put(300,0){\makebox(0,0){$\bullet$}}
\put(300,5){\makebox(0,0)[bc]{\small $\sqrt{s}=m_\xrm{\tiny L}$}}
\put(300,0){\vector(-1,-2){18}}
\put(293,-15){\makebox(0,0)[lt]{\small $\Delta E$}}
\put(281,-40){\makebox(0,0){$\bullet$}}
\put(286,-40){\makebox(0,0)[lc]{\small $K_\xrm{\tiny S}$ resonance pole}}
\put(5,20){\vector(0,1){20}}
\put(5,43){\makebox(0,0)[cl]{\small $\Imag{\sqrt{s}}$}}
\put(5,20){\vector(1,0){10}}
\put(18,20){\makebox(0,0)[cl]{\small $\Real{\sqrt{s}}$}}
\end{picture}
\end{center}
\caption[]{\small $K_{L}$ bound-state pole embedded in the continuum, and
$K_{S}$ resonance pole in the second Riemann sheet.}
\label{KSKLpoles}
\end{figure}
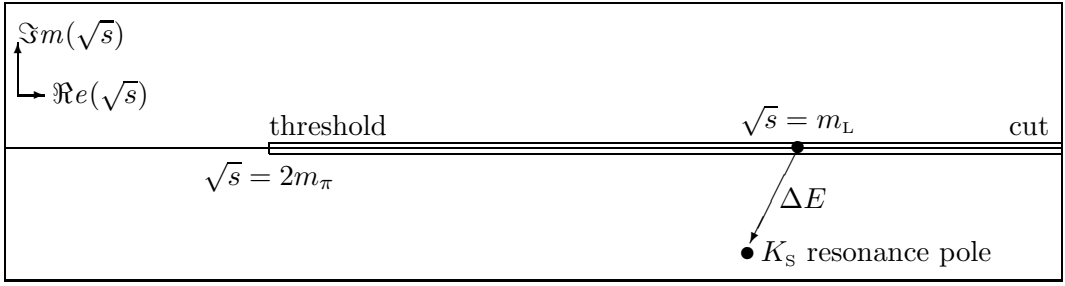

From expression~(\ref{cotgdS}) it is easy to perturbatively extract
the real and imaginary part of the resonance pole position in the
complex-energy plane, i.e.,
\begin{equation}
E_\xrm{\footnotesize pole}\;\approx\; E_{0}\; +\;\Delta E
\;\;,\;\;\xrm{with}\;\;\;\;
\Real{\Delta E}\; =\; R(s)\;\abs{{\cal F}_{0}}^{2}
\;\;\;\xrm{and}\;\;\;
\Imag{\Delta E}\; =\; I(s)\;\abs{{\cal F}_{0}}^{2}
\;\;\; .
\label{RIKS}
\end{equation}
The position of the $K_{S}$ resonance pole
with respect to the $K_{L}$ mass is shown in
Fig.~\ref{KSKLpoles}.
Since the coupling of the neutral kaons to the two-pion continuum is
extremely small,
the arrow pointing from $\sqrt{s}=m_{L}$ to the $K_{S}$ resonance pole position
represents the pole trajectory for increasing intensity of the transition
potential $V_{1}$, too.

For the ratio of the $K_{S}$ decay width $\Gamma_{S}$,
equalling twice the imaginary part (\ref{RIKS})
of the resonance pole position in the complex-energy plane,
and the neutral-kaon mass difference, which equals the real shift (\ref{RIKS})
of the pole with respect to $E_{0}=m_{L}$, we read from formula~(\ref{cotgdS})
the result
\begin{equation}
\fnd{\fsml{1}{2}\Gamma_{S}}
{m_{S}-m_{L}}\;\approx\;
\fnd{I(s)}{R(s)}
\;\;\; .
\label{ratiodef}
\end{equation}
From Ref.~\cite{HEPPH0304105} we moreover learn that, in the case of $S$-wave
scattering, one has
\begin{equation}
\fnd{I(s)}{R(s)}\; =\;\fnd{j_{0}(ka)}{n_{0}(ka)}\; =\;
-\xrm{tg}(ka)
\;\;\; .
\label{ISdef}
\end{equation}
As a consequence of Eqs.~(\ref{cotgdS}), (\ref{ratiodef}), and
(\ref{ISdef}), we obtain an extremely simple relation for the
mass difference in the neutral kaon system, the width of the
$K_{S}$ meson, the two-pion momentum $k$
and the strong-interaction radius $a$, reading
\begin{equation}
\fnd{\fsml{1}{2}\Gamma_{S}}
{m_{L}-m_{S}}\;\approx\;
\xrm{tg}(ka)
\;\;\; .
\label{ratio}
\end{equation}
Substitution of $k=0.206$ GeV and $a=3.2$ GeV$^{-1}$ gives us then the result
\begin{equation}
\fnd{\fsml{1}{2}\Gamma_{S}}
{m_{L}-m_{S}}\;\approx\; 1.3
\;\;\;\;\; (\xrm{Experiment:}\;\; 1.06 \mbox{\cite{PRD66p010001}})
\;\;\; .
\label{rationum}
\end{equation}

\noindent
Similar conclusions can be found in
Refs.~\cite{PLB238p20,PLB148p205,PLB152p251,PLB135p481,HEPPH0208118},
where long-distance effects have been studied in more detail.

\section{The \bm{J^{P}\!=\!1^{+}} \bm{c\bar{s}} states}

In the case of the $J^{P}\!=\!1^{+}$ $c\bar{s}$ states, we deal with two
distinct systems, $^{3\!}P_{1}$ and $^{1\!}P_{1}$, which, as far as confinement
is concerned, are degenerate when ignoring possible spin-orbit effects. Both
states couple strongly to $D^{\ast}\!K$, with threshold at about 2.501 GeV.

For the neutral kaon system, we could straightforwardly assume that
the transition potentials $V_{1}$ and $V_{2}$ in Eq.~(\ref{dynt}) are equal,
because of particle-antiparticle symmetry.
But for the $1^{+}$ $c\bar{s}$ systems we do not have such simple arguments.
Nevertheless, if $V_{1}$ and $V_{2}$ are proportional,
we still are in a situation comparable to the one discussed
in the previous section, though now for strong interactions.

In Ref.~\cite{ZPC21p291} it is shown that the spatial form of the
transition potential mainly depends on the orbital angular momenta of the
$c\bar{s}$ and $D^{\ast}\!K$ channels, whereas the relative intensities
follow from recoupling coefficients.
For transitions to vector+pseudoscalar we find that the intensity for
$^{3\!}P_{1}$ is twice as large as for $^{1\!}P_{1}$.
Consequently, by the use of Eqs.~(\ref{dyn3}) and (\ref{dyn12}),
we conclude that the $J^{P}\!=\!1^{+}$ $c\bar{s}$ bound state embedded in the
$D^{\ast}\!K$ continuum consists of a mixture of 33\% $^{3\!}P_{1}$ and 67\%
$^{1\!}P_{1}$, whereas its supposed resonance partner has more or less the
opposite content, but additionally with some $D^{\ast}\!K$ admixture.

In Sec.~(\ref{neutralK}) we have used Eq.~(\ref{dyn12})
for the description of $\pi\pi$ scattering in the presence of the
weak coupling to the neutral kaon system. Here, we assume that
Eq.~(\ref{dyn12}) is also suited to describe $D^{\ast}\!K$ scattering in the
presence of an infinity of $c\bar{s}$ confinement
states. In our unitarised model, this just implies substituting the
effective nonstrange quark mass by the charmed quark mass, the pion masses
by the $D^{\ast}$ and $K$ masses, and changing some of the quantum numbers.
The scattering phase shift is given by an expression similar to the one
shown in Eq.~(\ref{cotgdS}), but, since the interaction is not weak now,
also higher radial excitations of the $c\bar{s}$ confinement spectrum
must be included.  Thus, we use the general expression
\cite{HEPPH0306155,EPJC22p493}

\begin{equation}
\xrm{cotg}\left(\delta (s)\right)\; =\;
\fnd{n_{0}\left( pa\right)}{j_{0}\left( pa\right)}\; -\;
\left[ 2\lambda^{2}\mu pa
j^{2}_{0}\left( pa\right)\;
\sum_{n=0}^{\infty}\fnd{\abs{{\cal F}_{n}^{\: c\bar{s}}(a)}^{2}}
{\sqrt{s}-E_{n}}
\right]^{-1}
\;\;\; .
\label{Swave}
\end{equation}

Expression (\ref{Swave}) is not only valid above threshold, where for small
coupling ($\lambda\ll 1$) it generates Breit-Wigner-like resonance
structures in the scattering cross section around each of the energy
eigenvalues $E_{n}$ ($n=0$, 1, 2, $\dots$),
but it is also valid below threshold.
Resonances are characterized by complex-energy poles which are
given by solutions of $\xrm{cotg}\left(\delta (s)\right) =i$.
When solutions of $\xrm{cotg}\left(\delta (s)\right) =i$ come on the
real-energy axis below threshold, then they describe the real
energy eigenvalues for the bound state solutions of equation \ref{dyn12}.
For small values of $\lambda$ the latter poles are found in the
proximity of those $E_{n}$ ($n=0$, 1, 2, $\dots$) which come below
threshold.

The non-relativistic form of the Flatt\'{e} formula \cite{Flatte} can be
obtained from formula (\ref{Swave}) in the case of overlapping resonances.
But, then its analytic continuation to below threshold is lost.

From expression (\ref{Swave}) one can study the behavior of resonance poles,
just being solutions of $\xrm{cotg}\left(\delta (s)\right) =i$,
in function of variations of the model parameters
(see {\it e.g.} Ref~\cite{HEPPH0304105}).
Their passage from above threshold, where resonance poles exist in the
the second Riemann sheet, to below threshold, where poles are expected to
reside on the real-energy axis in the first Riemann sheet, is smooth and
without any difficulties.
For $S$-wave scattering the ground-state pole hits the real-energy axis below
threshold, then, for increasing values of $\lambda$,
moves along the real-energy axis towards threshold where it changes
Riemann sheet, turns 180 degrees and starts moving towards smaller values
of energy.

Here, similarly to the procedure outlined in
Ref.~\cite{HEPPH0306155,EPJC22p493},
we approximate the full sum over all $c\bar{s}$ confinement states
by the two nearest states, that is, the ground state at $E_{0}$ and the first
radial excitation at $E_{1}$, plus a rest term. The latter is scaled to 1 by
absorbing its value into the coupling constant $\lambda$, yielding

\begin{equation}
\sum_{n=0}^{\infty}\fnd{\abs{{\cal F}_{n}^{\: c\bar{s}}(a)}^{2}}
{\sqrt{s}-E_{n}}
\longrightarrow\;
\fnd{0.5}{\sqrt{s}-E_{0}}\; +\; \fnd{0.2}{\sqrt{s}-E_{1}}\; -\; 1
\;\;\;\xrm{GeV}^{\: 2}
\;\;\; .
\label{approx}
\end{equation}

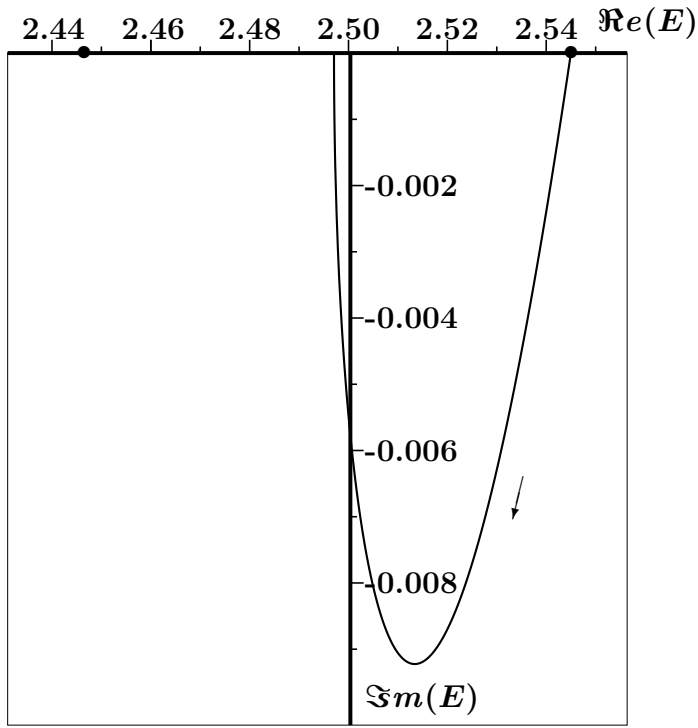
\begin{figure}[ht]
\normalsize
\begin{center}
\begin{picture}(283.46,293.46)(0.00,0.00)
\put(17.23,259.94){\makebox(0,0)[bc]{\bf 2.44}}
\put(55.07,259.94){\makebox(0,0)[bc]{\bf 2.46}}
\put(91.54,259.94){\makebox(0,0)[bc]{\bf 2.48}}
\put(128.93,259.94){\makebox(0,0)[bc]{\bf 2.50}}
\put(166.32,259.94){\makebox(0,0)[bc]{\bf 2.52}}
\put(203.71,259.94){\makebox(0,0)[bc]{\bf 2.54}}
\put(135.16,53.82){\makebox(0,0)[lc]{\bf -0.008}}
\put(135.16,103.97){\makebox(0,0)[lc]{\bf -0.006}}
\put(135.16,154.12){\makebox(0,0)[lc]{\bf -0.004}}
\put(135.16,204.27){\makebox(0,0)[lc]{\bf -0.002}}
\put(223,259.94){\makebox(0,0)[bl]{\bm{\Real{E}}}}
\put(135.16,4.01){\makebox(0,0)[bl]{\bm{\Imag{E}}}}
\put(28.89,254.42){\makebox(0,0){$\bullet$}}
\put(213,254.42){\makebox(0,0){$\bullet$}}
\put(195,94){\vector(-1,-4){4}}
\end{picture}
\end{center}
\normalsize
\caption[]{Trajectory of the ground-state pole in the $D^{\ast}\!K$
scattering amplitude, as a function of the coupling constant $\lambda$,
defined in Eq.~(\ref{Swave}), and which parametrises
the intensity of the transitions between the pure $c\bar{s}$ system and the
$D^{\ast}\!K$ continuum (see Eq.~\ref{dyn12}).
The arrow indicates how the pole moves when $\lambda$ is increased.
For vanishing coupling, as well as for the confined system described by
Eq.~(\ref{dyn3}), the pole is on the real $E$ axis. For large coupling we find
the pole on the real $E$ axis below threshold. Units are in GeV.}
\label{DsPpole}
\end{figure}

\noindent
We determine the ground-state mass of the uncoupled system~(\ref{dyn3})
from the model parameters
($m_{s}=0.508$, $m_{c}=1.562$ and $\omega =0.19$ GeV)
given in Ref.~\cite{PRD27p1527}, yielding $E_{0}=$2.545 GeV.
This also means that the bound state embedded in
the $D^{\ast}\!K$ continuum has exactly the mass $E_{0}$, only 9 MeV away from
the experimental \dsn\ mass \cite{PRD66p010001}. The first radial excitation
lies $2\omega=380$ MeV higher.

In Fig.~\ref{DsPpole} we study how the position of the ground-state
singularity in the scattering amplitude for Eq.~(\ref{dyn12})
varies, as the intensity of the transition potential $V_{1}$ is changed.
In Nature one can only ``measure'' the pole position for one particular value
of this intensity, given by strong interactions.
Nevertheless, it is very illustrative to study alternative values.
In particular, we notice that for larger intensities of the transition
mechanism the pole comes out on the real $E$ axis, below the $D^{\ast}\!K$
threshold. Had we started from the perturbative formula (\ref{cotgdS}),
where $R(s)$ and $I(s)$ are proportional to $\lambda^{2}$,
we would have started out as in Fig.~\ref{KSKLpoles}, and never
returned to the real $E$ axis for increasing values of the transition
intensity. The trajectory in Fig.~\ref{DsPpole} is highly nonperturbative,
which can only be achieved when all orders \cite{HEPPH0304105} are accounted
for.  The masses obtained for the model's values of $\lambda$ are indicated by
$\bullet$ in Fig.~\ref{DsPpole}. For the $D^{\ast}_{sJ}(2463)$ we read from the
figure $m=2.446$ GeV, so, after all, this state is not a resonance as one would
naively (i.e., perturbatively) expect from the coupled set of
equations (\ref{dyn12}), in agreement with experiment. As mentioned above,
for the $D_{s1}(2536)$ we obtain 2.545 GeV from pure confinement.

When the transition potentials $V_{1}$ and $V_{2}$ in Eq.~(\ref{dynt})
are not proportional, one has no simple diagonalisation, since
commutators with the Hamiltonians will spoil the simplicity.
Also, if $H_{1}$ and $H_{2}$ in Eq.~(\ref{dynt}) are not equal,
diagonalisation will not lead to completely decoupled systems.
However, from experiment we learn that the $D^{\ast}\!K$ width of the
$D_{s1}$(2536) is small (less than 2.3 MeV \cite{PRD66p010001}), implying that
the bulk of the interaction indeed stems from the unitarisation mechanism.

\section{The \bm{J^{P}\!=\!1^{+}} \bm{c\bar{n}} states}
\label{cbarn}

In order to describe the ground states of the $J^{P}\!=\!1^{+}$ $c\bar{n}$
spectrum, we use the same Eqs.~(\ref{cotgdS}) and (\ref{RIKS}) as in the
previous case.  Also here, we determine the ground-state mass of the uncoupled
system~(\ref{dyn3}) from the model parameters
($m_{u,d}=0.406$, $m_{c}=1.562$ and $\omega =0.19$ GeV)
in Ref.~\cite{PRD27p1527},
which now yields $E_{0}=$2.443 GeV. Hence, the $J^{P}\!=\!1^{+}$ $c\bar{n}$
bound state embedded in the $D^{\ast}\!\pi$ continuum comes out, in our model,
at 2.443 GeV, some 20 MeV above the experimental \cite{PRD66p010001,BELLE02}
\dn\ mass.  The first radial excitation lies, as before, $2\omega =380$ MeV
higher.
The remaining parameters $a$ and $\lambda$ are kept the same as in the
previous case, yet scaled with the reduced constituent $q\bar{q}$ mass
$\mu_{qq}$, in order to guarantee flavor invariance of strong interactions,
{\it i.e.}

\begin{equation}
a_{xy}\,\sqrt{\mu_{xy}}\; =\;\xrm{constant}
\;\;\;\;\xrm{and}\;\;\;\;
\lambda_{xy}\,\sqrt{\mu_{xy}}\; =\;\xrm{constant}
\;\;\;\; ,
\label{flavorinvariance}
\end{equation}
where $x$ and $y$ represent the two flavors involved.
The constants of formula (\ref{flavorinvariance}) are fixed by
\cite{HEPPH0110156} $a_{us}=3.2$ GeV$^{-1}$ and $\lambda_{us}=0.75$
GeV$^{-3/2}$ for $S$ wave.

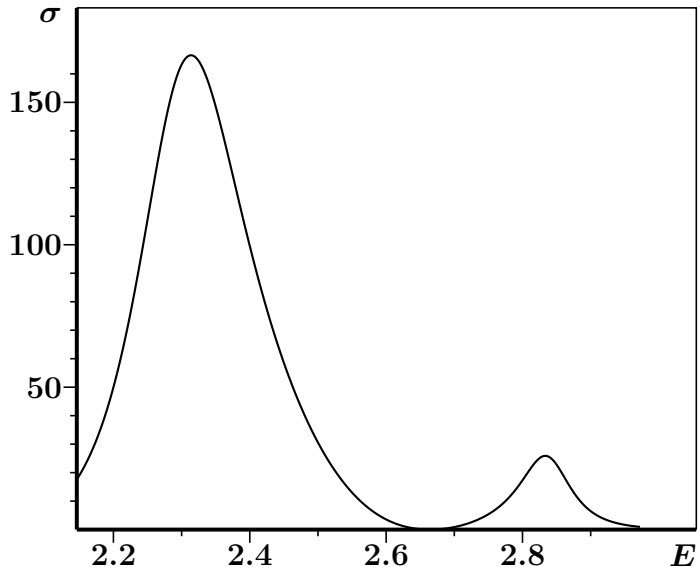
\begin{figure}[ht]
\normalsize
\begin{center}
\begin{picture}(283.46,236.77)(-50.00,-30.00)
\put(13.86,-5.52){\makebox(0,0)[tc]{\bf 2.2}}
\put(65.43,-5.52){\makebox(0,0)[tc]{\bf 2.4}}
\put(117.01,-5.52){\makebox(0,0)[tc]{\bf 2.6}}
\put(168.58,-5.52){\makebox(0,0)[tc]{\bf 2.8}}
\put(-5.52,53.91){\makebox(0,0)[rc]{\bf 50}}
\put(-5.52,107.82){\makebox(0,0)[rc]{\bf 100}}
\put(-5.52,161.73){\makebox(0,0)[rc]{\bf 150}}
\put(234.34,-5.52){\makebox(0,0)[tr]{$\bm{E}$}}
\put(-5.52,197.51){\makebox(0,0)[tr]{$\bm{\sigma}$}}
\end{picture}
\end{center}
\normalsize
\caption[]{Model result for the cross section in units of GeV$^{-2}$
for elastic $S$-wave $D^{\ast}\!\pi$ scattering, as a function of the
total invariant mass $E$ in units of GeV.}
\label{DnXs}
\end{figure}

The resulting cross section is given in Fig.~\ref{DnXs}.
Our peak shows up somewhat above 2.3 GeV, whereas the width of our \dn\ is
about 200 MeV. The corresponding experimental values are
$2400\pm 30\pm 20$ MeV \cite{BELLE02} ($2461^{+48}_{-42}$ MeV \cite{CLEO01}),
and $380\pm 100\pm 100$ MeV \cite{BELLE02} ($290^{+110}_{-90}$ MeV
\cite{CLEO01}), respectively.
In Fig.~\ref{DnXs} we can further observe that the first radial excitation of
the system of equations~(\ref{dyn12}), for $J^{P}\!=\!1^{+}$ $c\bar{n}$, comes
out more than 500 MeV higher than the ground-state resonance. Nevertheless, in
expression~(\ref{RIKS}) we always used $E_{1}-E_{0}=380$ MeV \cite{PRD27p1527}.
Such a highly non-perturbative behaviour is inherent in the unitarisation
procedure leading to formula~(\ref{cotgdS}).

The bound state $D_{1}$(2420) embedded in the $D^{\ast}\!\pi$ continuum has
zero width in our model, and so is invisible in the cross section of
Fig.~\ref{DnXs}. On the other hand, experiment finds $18.9^{+4.6}_{-3.5}$ MeV
\cite{PRD66p010001} and $26.7\pm 3.1\pm 2.2$ \cite{BELLE02} MeV for the full
width, which is nonetheless very small as compared to the available phase
space. Consequently, also in this case our assumption in Eq.~(\ref{V2propV1})
appears to be reasonable.

The relative intensities $\alpha$ and $\beta$ in Eq.~(\ref{V2propV1})
are the same as in the $c\bar{s}$ case. Hence, the relative content of the
$D_{1}$(2420) is again 33\% $^{3\!}P_{1}$ and 67\% $^{1\!}P_{1}$.
The $D_{1}$(2400) roughly has the opposite mixture, plus a $D^{\ast}\!\pi$
component.

\begin{figure}[ht]
\normalsize
\begin{center}
\begin{picture}(283.46,293.46)(-50.00,0.00)
\put(42.24,259.94){\makebox(0,0)[bc]{\bf 2.2}}
\put(112.53,259.94){\makebox(0,0)[bc]{\bf 2.3}}
\put(182.82,259.94){\makebox(0,0)[bc]{\bf 2.4}}
\put(-1.05,40.85){\makebox(0,0)[rc]{\bf -0.10}}
\put(-1.05,83.56){\makebox(0,0)[rc]{\bf -0.08}}
\put(-1.05,126.27){\makebox(0,0)[rc]{\bf -0.06}}
\put(-1.05,168.99){\makebox(0,0)[rc]{\bf -0.04}}
\put(-1.05,211.70){\makebox(0,0)[rc]{\bf -0.02}}
\put(239,259.94){\makebox(0,0)[br]{\bm{\Real{E}}}}
\put(-1.05,4.01){\makebox(0,0)[br]{\bm{\Imag{E}}}}
\put(190,120){\vector(-1,-4){4}}
\put(213,254){\makebox(0,0){$\bullet$}}
\put(108.68,30.96){\makebox(0,0){$\bullet$}}
\normalsize
\end{picture}
\end{center}
\normalsize
\caption[]{Trajectory of the ground-state pole in the $D^{\ast}\!\pi$
scattering amplitude, as a function of the coupling constant $\lambda$.
The arrow indicates how the pole moves when $\lambda$ is increased.
For vanishing coupling, as well as for the confined system described by
Eq.~(\ref{dyn3}), here representing the $D_{1}$(2420),
the pole is on the real $E$ axis. The pole postion for Eq.~(\ref{dyn12}), which
describes the $D_{1}$(2400), is indicated by $\bullet$ for the physical value
of $\lambda$. Units are in GeV.}
\label{DnPpole}
\end{figure}
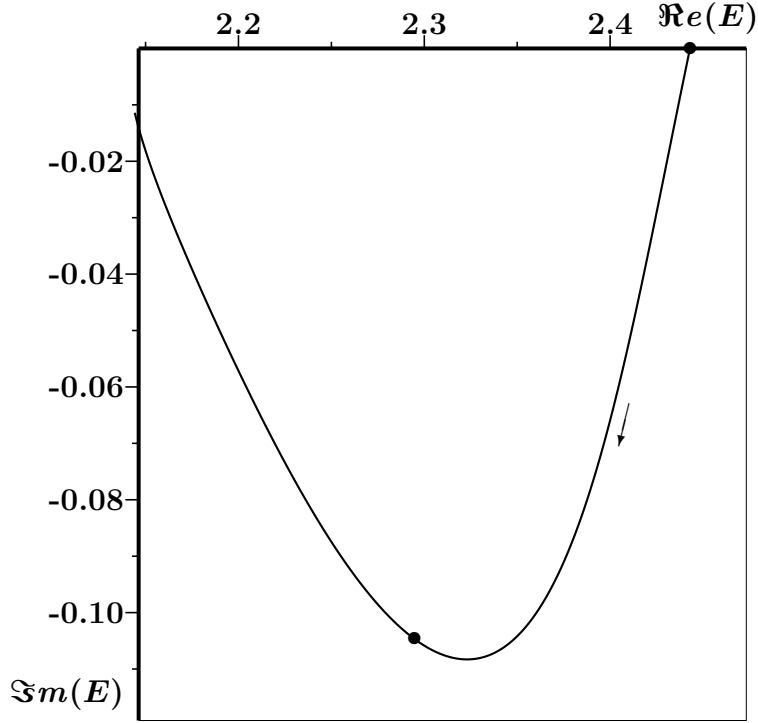

In Fig.~\ref{DnPpole} we show how the scattering pole moves in the
complex-energy plane when $\lambda$ is varied, which is the pole associated
with the cross section of Fig.~\ref{DnXs}.

\section{\bm{J^{P}\!=\!1^{+}} systems containing one \bm{b} quark}

For $J^{P}\!=\!1^{+}$ systems containing one $b$ quark we may repeat the above
described procedure.
We determine the ground-state masses of the uncoupled systems~(\ref{dyn3})
from the model parameters
($m_{u,d}=0.406$, $m_{s}=0.508$, $m_{c}=1.562$, $m_{b}=4.724$
and $\omega =0.19$ GeV)
given in Ref.~\cite{PRD27p1527}.
The remaining parameters $a$ and $\lambda$ are kept the same as for the
case of the $D$ mesons.
This results in 5.707, 5.809 and 6.863 GeV for respectively $\bar{b}u/d$,
$b\bar{s}$ and $\bar{b}c$.
The model's results for the partner states which couple to the continuum,
are collected in table (\ref{Bstates}).
Similar to the case of $D^{\ast}\!\pi$ (section~\ref{cbarn}), we find in
$B^{\ast}\!\pi$ a resonance structure above threshold, peak value at 5.6 GeV
and some 120 MeV wide.
The experimental result might be peaking at a slightly higher value
\cite{PLB425p215,PLB465p323,PRD64p072002,HEPEX0110049}, but the width
of this resonance is in good agreement with our result.
For the continuum bound state, which we find at 5.71 GeV,
the OPAL collaboration reports a narrow resonance at 5.74 GeV
\cite{HEPEX0110049},
whereas the CDF collaboration gives 5.719 GeV for the narrow $B_{1}$
resonance \cite{PRD64p072002}.
We also might compare with the prediction of Kalashnikova
and Nefediev, who obtain 5.716 and 5.741 GeV for the two $B_{1}$
states \cite{HEPPH0112330}.

It should be noticed from table~\ref{Bstates} that the $1^{+}$ $b\bar{s}$
partner state comes in our model at 5.73 GeV, which is in the same energy
region as the other two states.
However, the CDF collaboration reports a very small (3.7\%) contamination due
to $B_{s}$.

\begin{table}[ht]
\begin{center}
\begin{tabular}{|l||l|l|l|l|}
\hline & & continuum & & scattering\\
system & threshold & bound state & partner state & length\\ [.2cm]
& GeV & GeV & GeV & GeV$^{-1}$\\
\hline\hline & & & & \\
$\bar{b}u$ & 5.465 ($B^{\ast 0}\pi^{+}$) & 5.71 & $5.59-0.06i$ & -1.13\\ [.3cm]
$b\bar{s}$ & 5.819 ($B^{\ast -}K^{+}$) & 5.81 & 5.73 & 2.08\\ [.3cm]
$\bar{b}c$ & 7.190 ($B^{\ast 0}D^{+}$) & 6.86 & 6.83 & -31.8\\ \hline
\end{tabular}
\end{center}
\caption[]{Purely bound states and partners for
$J^{P}=1^{+}$ systems containing one $b$ quark.}
\label{Bstates}
\end{table}

Literally speaking, the $1^{+}$ $B_{s}$ and $B_{c}$ do not make part of the
combinations {\it continuum\--bound\--state/\-partner\--state}, since even
the purely bound states come below threshold.
However, the case of $b\bar{s}$ could be different experimentally,
given the error of some 10-30 MeV in our model predictions.
So, there might be a continuum bound state, slightly above the
$B^{\ast}K$ threshold.
But, the partner state is definitely below threshold and thus very narrow.

For $B_{c}$ both states are well below threshold.

\section{Summary and conclusions}
In the present paper, we have studied examples of mesonic systems that generate
bound or quasi-bound states in the continuum, within an unitarised quark-meson
model. In the neutral-kaon system, our approach provides a simple explanation
for the widths and the mass difference of the $K_L$ and $K_S$, which is also in
agreement with the conclusions of more sophisticated methods. Application to
the $J^{P}\!=\!1^{+}$ $c\bar{s}$ and $c\bar{n}$ axial-vector charmed mesons
accomplishes a simultaneous description of the established narrow \dsn\
and \dn\ states, as well as the recently observed very narrow \dsb,
assuming it indeed is a $1^{+}$ state, and broad \db, which is rather
problematic in standard quark models.

We predict the continuum bound states $D_{1}$(2420) and $D_{s1}(2536)$ to be
mixtures of 33\% $^{3\!}P_{1}$ and 67\% $^{1\!}P_{1}$,
which might be measured through radiative transitions \cite{HEPPH0305122}.
For their partners $D_{1}(2400)$ and $D^{\ast}_{sJ}(2463)$ we predict
more or less the opposite $^{2S+1}\!P_1$-state content, but additionally with
some $D^{\ast}\pi$ or $D^{\ast}\!K$ admixture, respectively.
This is in agreement with the mixtures proposed by Godfrey and Kokoski in
Ref.~\cite{PRD43p1679}, based on the flux-tube-breaking mechanism.
Their $\theta =-38^{\circ}$ for $c\bar{s}$ amounts to 38\% $^{3\!}P_{1}$ and
62\% $^{1\!}P_{1}$ regarding the state {\it degenerate} \/with
$^{3\!}P_{2}$, and the opposite content for the state {\it degenerate} \/with
$^{3\!}P_{0}$. The former mixture corresponds to the continuum bound state
$D_{s1}(2536)$, which is indeed approximately degenerate with the $^{3\!}P_{2}$
in our model, since this $J\!=\!2$ state is subject to only very small
unitarisation effects due to the $D$-wave $D^{\ast}\!K$ channel. However,
the latter mixture corresponds to the partner state $D^{\ast}_{s1}$ at
2.443 GeV, which is strongly influenced by the coupling to the $S$-wave
$D^{\ast}\!K$ channel, just as the likely $^3\!P_0$ $\;D^{\ast}_{sJ}(2317)^+$
\cite{HEPEX0304021} is drastically affected by the $S$-wave $DK$ threshold
\cite{HEPPH0305035}. Therefore, no approximate degeneracy holds for these two
charmed mesons. Concerning the $c\bar{u}$ states, Godfrey and Kokoski obtained
$\theta =-26^{\circ}$, resulting in 19\% $^{3\!}P_{1}$ and 81\% $^{1\!}P_{1}$,
which indicates that our assumption~(\ref{V2propV1}) is probably somewhat less
accurate in this case.

For $J^{P}\!=\!1^{+}$ quark-antiquark systems which contain one $b$ quark,
we obtain good results as far as we can compare to experiment.

In conclusion, we have once again demonstrated that unitarisation has to be
incorporated in realistic quark models of mesons and baryons, as it
constitutes the second most important interaction after confinement.
\vspace{1cm}

\noindent {\large\bf Acknowledgements} \\
This work was partly supported by the
{\it Funda\c{c}\~{a}o para a Ci\^{e}ncia e a Tecnologia}
\/of the {\it Minist\'{e}rio da
Ci\^{e}ncia e do Ensino Superior} \/of Portugal,
under contract number
POCTI/\-FNU/\-49555/\-2002.


\begin{thebibliography}{36}
\bibitem{PZ30p465}
J.~von~Neumann and E.~Wigner,
Phys. Z. {\bf 30}, 465 (1929).

\bibitem{PRL90p013001}
Lorenz~S.~Cederbaum, Ronald~S.~Friedman, Victor~M.~Ryaboy and Nimrod Moiseyev,
Phys.\ Rev.\ Lett.\  {\bf 90}, 013001 (2003).

\bibitem{SAMARA97p435}
D.~L.~Pursey and T.~A.~Weber,
prepared for {\it 12th Int. Workshop on High-Energy Physics and Quantum
Field theory (QFTHEP 97), Samara, Russia, 4-10 Sep 1997},
published in {\it Samara 1997}, QFTHEP'97, pp. 435-438.

\bibitem{GRQC9610018}
H.~C.~Rosu and J.~Socorro,
Nuovo Cim.\ B {\bf 113}, 677 (1998)
[arXiv:gr-qc/9610018].

\bibitem{QUANTPH9511022}
A.~Khelashvili and N.~Kiknadze,
J.\ Phys.\ A {\bf 29}, 3209 (1996)
[arXiv:quant-ph/9511022].

\bibitem{HEPPH9305336}
J.~Pappademos, U.~Sukhatme and A.~Pagnamenta,
Phys.\ Rev.\ A {\bf 48}, 3525 (1993)
[arXiv:hep-ph/9305336].

\bibitem{HEPPH0201006}
Eef van Beveren and George Rupp,
in Proc. Workshop {\it Recent Developments in Particle
and Nuclear Physics, April 30, 2001, Coimbra (Portugal)},
(Universidade de Coimbra, 2003) ISBN 972-95630-3-9, pages 1--16,
[arXiv:hep-ph/0201006].

\bibitem{PRD66p010001}
K.~Hagiwara {\it et al.} \/[Particle Data Group Collaboration],
Phys.\ Rev.\ D {\bf 66}, 010001 (2002).

\bibitem{HEPEX0305100}
D.~Besson  [CLEO Collaboration],
arXiv:hep-ex/0305100.

\bibitem{HEPEX0305017}
D.~Besson {\it et al.}  [CLEO Collaboration],
to appear in Proc. {\it 8th Conference on the Intersections of
Particle and Nuclear Physics (CIPANP 2003), New York, 19-24 May 2003},
arXiv:hep-ex/0305017.

\bibitem{BELLE02}
K.~Abe {\it et al.} [BELLE Collaboration],
BELLE-CONF-02-35, Cont.\ paper for
the {\it 31st Int.\ Conf.\ on High Energy Physics
(ICHEP 2002), Amsterdam, The Netherlands, 24--31 Jul 2002},
session 8,
{\it Heavy quark mesons and baryons (incl. lattice calculations)},
paper ABS724.

\bibitem{CLEO01}
S.~Anderson {\it et al.} [CLEO Collaboration],
Conference report CLEO-CONF-99-6, (1999).

\bibitem{ZPC21p291}
E.~van Beveren,
Z.\ Phys.\ {\bf C21} (1984) 291.

\bibitem{HEPPH0306155}
Eef van Beveren and George Rupp,
Talk given at
{\it The 25th annual Montreal-Rochester-Syracuse-Toronto
Conference on High-Energy Physics}, Joefest,
{\it in the honor of the 65th birthday of Joseph Schechter,
May 13 - 15, 2003, Syracuse (NY)},
to appear in AIP Conf.\ Proc.\ (2003),
arXiv:hep-ph/0306155.

\bibitem{CPC27p377}
C.~Dullemond, G.~Rupp, T.~A.~Rijken, and E.~van Beveren,
Comput.\ Phys.\ Commun.\ {\bf 27} (1982) 377.

\bibitem{PRD27p1527}
E.~van Beveren, G.~Rupp, T.~A.~Rijken, and C.~Dullemond,
Phys.\ Rev.\ D {\bf 27}, 1527 (1983).

\bibitem{PRD21p772}
E.~van Beveren, C.~Dullemond, and G.~Rupp,
Phys.\ Rev.\ {\bf D21} (1980) 772
[Erratum-ibid.\ {\bf D22} (1980) 787].

\bibitem{EPJC22p493}
Eef van Beveren and George Rupp,
Eur.\ Phys.\ J.\ C {\bf 22}, 493 (2001)
[arXiv:hep-ex/0106077].

\bibitem{HEPPH0110156}
Eef van Beveren and George Rupp,
AIP Conf.\ Proc.\  {\bf 619}, 209 (2002)
[arXiv:hep-ph/0110156].

\bibitem{MPLA14p1273}
M.~D.~Scadron,
Mod.\ Phys.\ Lett.\ A {\bf 14}, 1273 (1999)
[arXiv:hep-ph/9910244].

\bibitem{HEPPH0304105}
E.~van Beveren and G.~Rupp,
Int. J. Theor. Phys. Group Theor. Nonlin. Opt., in press (2003)
arXiv:hep-ph/0304105.

\bibitem{PLB238p20}
G.~E.~Brown, J.~W.~Durso, M.~B.~Johnson and J.~Speth,
Phys.\ Lett.\ B {\bf 238}, 20 (1990).

\bibitem{PLB148p205}
I.~I.~Bigi and A.~I.~Sanda,
Phys.\ Lett.\ B {\bf 148}, 205 (1984).

\bibitem{PLB152p251}
P.~Cea and G.~Nardulli,
Phys.\ Lett.\ B {\bf 152}, 251 (1985).

\bibitem{PLB135p481}
John F.~Donoghue, Eugene Golowich and Barry R.~Holstein,
Phys.\ Lett.\ B {\bf 135}, 481 (1984).

\bibitem{HEPPH0208118}
J.~Lowe and M.~D.~Scadron,
arXiv:hep-ph/0208118.

\bibitem{Flatte}
Stanley M.~Flatt\'{e},
Phys.\ Lett.\ B {\bf 63}, 224 (1976).

\bibitem{PLB425p215}
R.~Barate {\it et al.}  [ALEPH Collaboration],
Phys.\ Lett.\ B {\bf 425}, 215 (1998).

\bibitem{PLB465p323}
M.~Acciarri {\it et al.}  [L3 Collaboration],
Phys.\ Lett.\ B {\bf 465}, 323 (1999)
[arXiv:hep-ex/9909018].

\bibitem{PRD64p072002}
T.~Affolder {\it et al.}  [CDF Collaboration],
Phys.\ Rev.\ D {\bf 64}, 072002 (2001).

\bibitem{HEPEX0110049}
K.~Harder,
arXiv:hep-ex/0110049.

\bibitem{HEPPH0112330}
Y.~S.~Kalashnikova and A.~V.~Nefediev,
Phys.\ Lett.\ B {\bf 530}, 117 (2002)
[arXiv:hep-ph/0112330].

\bibitem{HEPPH0305122}
Stephen~Godfrey,
arXiv:hep-ph/0305122.

\bibitem{PRD43p1679}
Stephen~Godfrey and Richard~Kokoski,
Phys.\ Rev.\ D {\bf 43}, 1679 (1991).

\bibitem{HEPEX0304021}
B.~Aubert {\it et al.}  [BABAR Collaboration],
arXiv:hep-ex/0304021.

\bibitem{HEPPH0305035}
Eef~van Beveren and George~Rupp,
arXiv:hep-ph/0305035.
\end{thebibliography}
\end{document}